\documentclass[aps, preprint, groupedaddress, floatfix]{revtex4}

\usepackage{epsfig}
\usepackage{graphicx}
\usepackage{amsmath}
\usepackage{rotating}
\usepackage{color}
\usepackage{multirow}

\begin{document}

\title{Antisite defects at oxide interfaces}

\author{Hanghui~Chen$^{1,2}$ and Andrew Millis$^{1}$}

\affiliation{
 $^1$Department of Physics, Columbia University, New York, NY, 10027, USA\\
 $^2$Department of Applied Physics and Applied Math, Columbia University, New York, NY, 10027,
 USA\\}
\date{\today}

\begin{abstract}
We use \textit{ab initio} calculations to estimate formation energies
of cation (transition metal) antisite defects at oxide interfaces and
to understand the basic physical effects that drive or suppress the
formation of these defects. Antisite defects are found to be favored
in systems with substantial charge transfer across the interface,
while Jahn-Teller distortions and itinerant ferromagnetism can prevent
antisite defects and help stabilize atomically sharp interfaces. Our
results enable identification of classes of systems that may be more
and less susceptible to the formation of antisite defects and motivate
experimental studies and further theoretical calculations to elucidate
the local structure and stability of oxide interface systems.
\end{abstract}

\maketitle

\section{Introduction}

The remarkable electronic properties of transition metal oxides,
including high transition temperature
superconductivity~\cite{Lee-RMP-2006}, colossal
magnetoresistance~\cite{Salamon-RevModPhys-2001} and metal-insulator
transitions~\cite{Imada-RMP-1998} make them of fundamental importance
for condensed matter physics. Interest has significantly increased
following the fabrication of atomic-precision heterointerfaces which
bring together different transition metal oxides with different bulk
properties
\cite{Hwang-Nature-2002,Hwang-Nature-2004,Chakhalian-NatPhys-2006,Chakhalian-Science-2007,Yoshimatsu-Science-2011,Monkman-NatMat-2012,May-NatMat-2009}. This
ability to control materials at the atomic scale holds out the promise
of creating systems with entirely new properties and
functionalities~\cite{Mannhart-MRS-2008,Chen-PRL-2013a,Chen-PRL-2013b,Chen-PRB-2014,Disa-PRL-2015,Cao-NatCommun-2016}.

Realizing these exciting possibilities requires atomically precise
interfaces. However, studies of interfaces separating simpler
semiconducting
materials~\cite{Bachelet-PRB-1983,Baraff-PRL-1985,Meyer-PRL-1984,Mattila-PRL-1995}
show that antisite defects (exchange of atoms across the interface)
may occur and can have crucial (and typically degrading) effects on
near-interface electronic properties. In particular, emergent
phenomena such as $d$-wave superconductivity and Weyl metal behavior are
typically sensitive to disorder, and clean samples are required for a
convincing observation~\cite{Chakhalian-RMP-2014, Adam-PNAS-2007, DasSarma-RMP-2011}.
Although important work, in
particular on defects at interfaces characterized by polar
discontinuities, has
appeared~\cite{Hwang-NatMat-2006,Chambers-SSR-2010,Freysoldt-RMP-2014,Yu-NatComm-2014},
the subject of antisite defects at oxide interfaces has received
relatively little attention.

In this paper we consider antisite defects at
$AM$O$_3$/$AM^\prime$O$_3$ interfaces separating different members of
the $AM$O$_3$ class of pseudocubic perovskite transition metal
oxides. In these materials the $A$-site is occupied by a lanthanide or
an alkali earth ion (we consider $A$=La or $A$=Sr) and the $M$-site is
occupied by a transition metal ion (we consider $M,M^\prime$ drawn
from the first transition metal row). We focus on the situation in
which the $A$-site is occupied by the same ion throughout and a change
in the $M$-site ion defines the interface so an antisite defect
corresponds to an exchange of $M$ and $M^\prime$ ions across the
interface~\cite{AA}. For all relevant combinations of $M$ and
$M^\prime$ we compute the defect formation energy, and then provide a
physical understanding of the results in terms of the relative
importance of charge transfer across the interface (leading to
octahedral volume disproportionation that favors defects) and
Jahn-Teller distortions (which inhibit defect formation). For metallic
systems, itinerant ferromagnetism emerges as an additional factor
inhibiting antisite defects.

The rest of this manuscript is organized as
follows. Section~\ref{Computational} outlines the methods used;
Section~\ref{Results} presents our principal results, namely energies
and local lattice structure for different antisite defect
combinations; Section~\ref{Distortions} gives an interpretation of the
results in terms of charge transfer, structural distortions and
itinerant ferromagentism; Section~\ref{Conclusion} is a summary and
conclusion.

\section{Computational Details \label{Computational}}

We perform density functional theory
calculations~\cite{Hohenberg-PR-1964, Kohn-PR-1965} within the \textit{ab
initio} supercell plane-wave approach~\cite{Payne-RMP-1992}, as
implemented in the Vienna Ab-initio Simulation Package
(VASP)~\cite{Kresse-PRB-1996}. We employ the Perdew, Burke and
Ernzerhof (PBE) parameterization~\cite{Perdew-PRL-1996} of the
generalized gradient approximation (GGA) to the Kohn-Sham potential
and projector augmented wave pseudopotentials~\cite{Blochl-PRB-1994,
  Kresse-PRB-1999}. The energy cutoff is 600 eV. We employ three
different types of simulation cells. For most of the calculations, we
use a $\sqrt{2}\times\sqrt{2}\times2$ supercell; to test the effects
of inter-defect interactions we use a $2\times2\times2$ supercell and
to understand the effects of interface-interface separations a
$\sqrt{2}\times\sqrt{2}\times 8$ supercell is used. A $8\times8\times6$
Monkhorst-Pack grid is used to sample the Brillouin zone of the
$\sqrt{2}\times\sqrt{2}\times2$ supercell. A $5\times5\times5$
Monkhorst-Pack grid is used to sample the Brillouin zone of the
$2\times2\times2$ supercell. A $8\times8\times2$ Monkhorst-Pack grid
is used to sample the Brillouin zone of the
$\sqrt{2}\times\sqrt{2}\times8$ supercell. Both cell and internal
coordinates are fully relaxed until each force component is smaller
than 10 meV/\AA~and the stress tensor is smaller than 10
kBar. Convergence of the key results was tested with a higher energy
cutoff and a denser $k$-point sampling and no significant changes were
found. Correlation effects on the $3d$ orbitals are included using the
VASP implementation of the rotationally invariant GGA+$U$
approximation introduced in Ref.~\cite{Lie-PRB-1995}. We use $U=$ 5.0
eV on the $d$ orbitals for all the transition metal ions
considered. For early transition metal ions ($M$=Ti, V and Cr), we
choose $J=$ 0.65 eV on the $d$ orbitals and for late transition metal
ions ($M$=Mn, Fe, Co and Ni), we choose $J=$ 1 eV on the $d$ orbitals.
In order to shift the empty La $4f$ states to higher energy, we also
use $U_{\textrm{La}}= 9.0$ eV on the $f$ orbital, following the value
used in previous work~\cite{RChen-PRB-2013}. While the GGA+$U$ method
is only an approximate solution of the correlated electron problem
posed by transition metal oxides, it is generally accepted as a robust
method that captures the important trends in ground state energy and is
computationally tractable, permitting surveys of wide ranges of
interfaces. The most significant errors are in dynamical quantities
that are not important for this work.

\section{Results \label{Results}}

\subsection{Inter-defect and interface-interface interactions}

\begin{figure}[t!]
\includegraphics[angle=0,width=0.8\columnwidth]{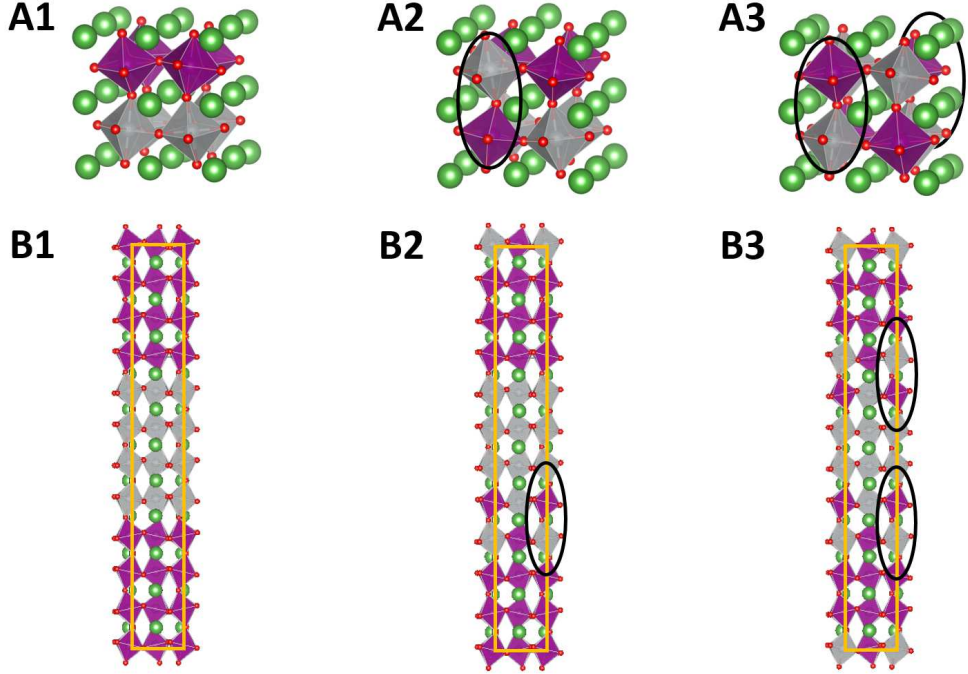}
\caption{\label{fig:supple-structure} Panel \textbf{A}: $2\times2\times2$ 
supercell. Panel \textbf{B}: $\sqrt{2}\times\sqrt{2}\times8$ supercell. 
Column \textbf{1}: ideal interfaces with no antisite defects. 
Column \textbf{2}: one antisite defect in the supercell.
Column \textbf{3}: two antisite defects in the supercell. Antisite defects 
are highlighted by the black ellipses.}
\end{figure}

\begin{figure}[t]
\includegraphics[angle=0,width=0.8\columnwidth]{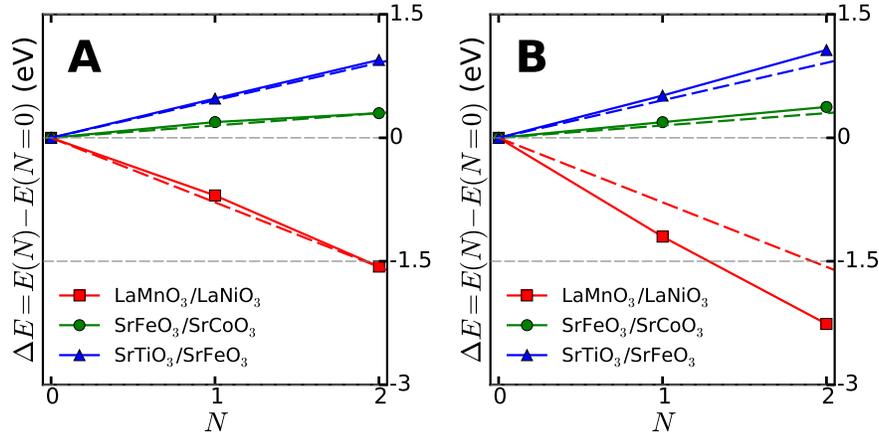}
\caption{\label{fig:large-cell} Solid symbols (solid lines guide to
  the eye): total energies (with respect to the total energy of ideal
  interfaces) of various antisite defect configurations 
  (see Fig.~\ref{fig:supple-structure}) calculated
  \textbf{A}) in a $2\times2\times2$ supercell (40-atom) and \textbf{B}) in a
  $\sqrt{2}\times\sqrt{2}\times8$ supercell (80-atom). Dashed lines:
  estimation of defect formation energies calculated in a
  $\sqrt{2}\times\sqrt{2}\times2$ supercell (20-atom). $N$ is the number of
  antisite defects in the supercell.}
\end{figure}

We define the defect formation energy as the difference between the
energy of systems with and without antisite defects. Computing the
energy of a single antisite defect at an isolated interface would
require an infinitely large computational cell. Practical calculations
employ finite supercells and therefore involve both a non-vanishing
defect density and a finite spacing between interfaces. To assess the
degree to which our finite supercell calculations are affected by
non-vanishing defect densities, we studied a $2\times 2\times2$
supercell (40 atom in total, see
Fig.~\ref{fig:supple-structure}\textbf{A1}) which can accommodate
either one or two antisite defects (see
Fig.~\ref{fig:supple-structure} \textbf{A2}
and~\ref{fig:supple-structure}\textbf{A3}). This corresponds to 25\%
or 50\% defect concentration per interface. We restricted attention to
ferromagnetic states to avoid issues of interplay between inter-defect
spacing and magnetic ordering wave vectors. Results for three
representative choices of $A$, $M$ and $M^\prime$ are shown in panel
{\bf A} of Fig.~\ref{fig:large-cell}. The symbols are the calculated
energy differences between a system with $N$ defects and a system with
none; the slopes of the dashed lines give the formation energies
estimated from the 50\% defect concentration calculations. We see that
using the higher defect concentration (50\%) provides a formation
energy which slightly overestimates that from the lower concentration
(25\%). The higher defect concentration (50\%) can be accommodated in
a smaller $\sqrt{2}\times\sqrt{2}\times 2$ simulation cell.

To assess the consequences of a finite distance between interfaces, we
study a $\sqrt{2}\times\sqrt{2}\times8$ supercell (80-atom in total)
in which the two interfaces are separated by four unit cells. We
consider three configurations: i) both interfaces are ideal
(Fig.~\ref{fig:supple-structure}\textbf{B1}); ii) one interface is
ideal and the other interface has antisite defects (50\% concentration
per interface) and iii) both interfaces have antisite defects (50\%
concentration per interface). We compare in panel {\bf B} of
Fig.~\ref{fig:large-cell} results obtained on a
$\sqrt{2}\times\sqrt{2}\times8$ supercell to results obtained on a
$\sqrt{2}\times\sqrt{2}\times2$ supercell. Isolating the
interfaces does not change the \textit{sign} of the energy difference
but does somewhat increase the magnitude. These results indicate that
a $\sqrt{2}\times\sqrt{2}\times 2$ supercell can be used as a
conservative estimator of antisite defect formation energy.

\subsection{Energetics}

\begin{figure}[t]
\includegraphics[angle=0,width=12cm]{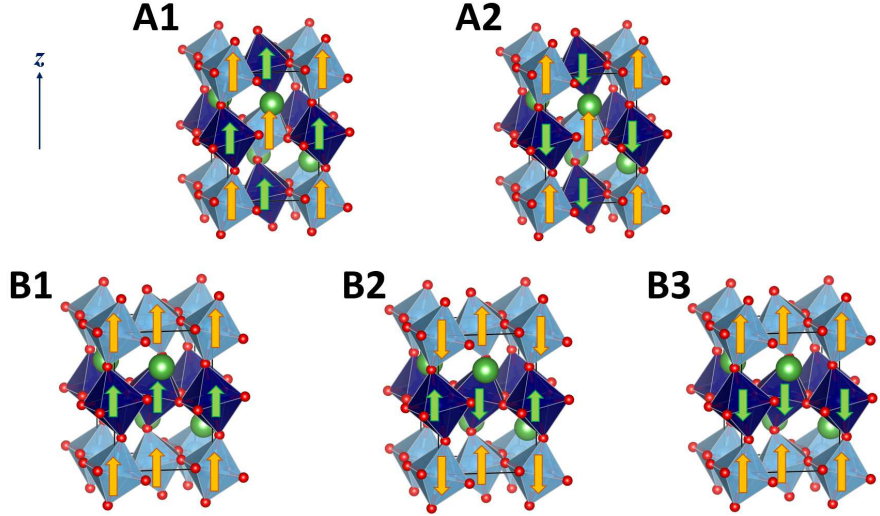}
\caption{\label{fig:structure-mag} Sketch of atomic and magnetic
  structures considered in this work. Panels \textbf{A}: rocksalt
  configuration.  Panels \textbf{B}: layered configuration.  Large orange
  balls denote $A$-site ions La or Sr; small red balls denote O ions;
  shaded octahedra denote octahedra containing transition metal $M$
  (darker shade) and $M^\prime$ (lighter shade) ions, respectively.
  \textbf{A1} and \textbf{B1} are ferromagnetic ordering; \textbf{A2}
  and \textbf{B2} are checkboard $G$-type antiferromagnetic ordering;
  \textbf{B3} are $A$-type antiferromagnetic ordering (spins are
  parallel in each layer and antiparallel between adjacent
  layers). Magnetic moments are schematically indicated by
  green/yellow arrows.}
\end{figure}

We use the $\sqrt{2}\times\sqrt{2}\times 2$ supercell to survey 21
La$M$O$_3$/La$M^\prime$O$_3$ interfaces ($M,M^\prime$ = Ti,V, Cr, Mn,
Fe, Co, Ni) and 15 Sr$M$O$_3$/Sr$M^\prime$O$_3$ interfaces
($M,M^\prime$ = Ti, V, Cr, Mn, Fe, Co). The
$\sqrt{2}\times\sqrt{2}\times 2$ simulation cell is illustrated in
Fig.~\ref{fig:structure-mag}; it consists of four perovskite primitive
cells (20 atoms in total) and is large enough to accommodate both
Jahn-Teller~\cite{Jahn-RSL-1937} and GdFeO$_3$
distortions~\cite{Gallasso} as well as N\'{e}el antiferromagnetic
ordering. The stacking direction of the layered structure is along the
$z$ axis.

\begin{figure*}[t]
%\includegraphics[angle=0,width=0.5\textwidth]{raw-fit}
%\hspace{-1.5cm}
\includegraphics[angle=0,width=\columnwidth]{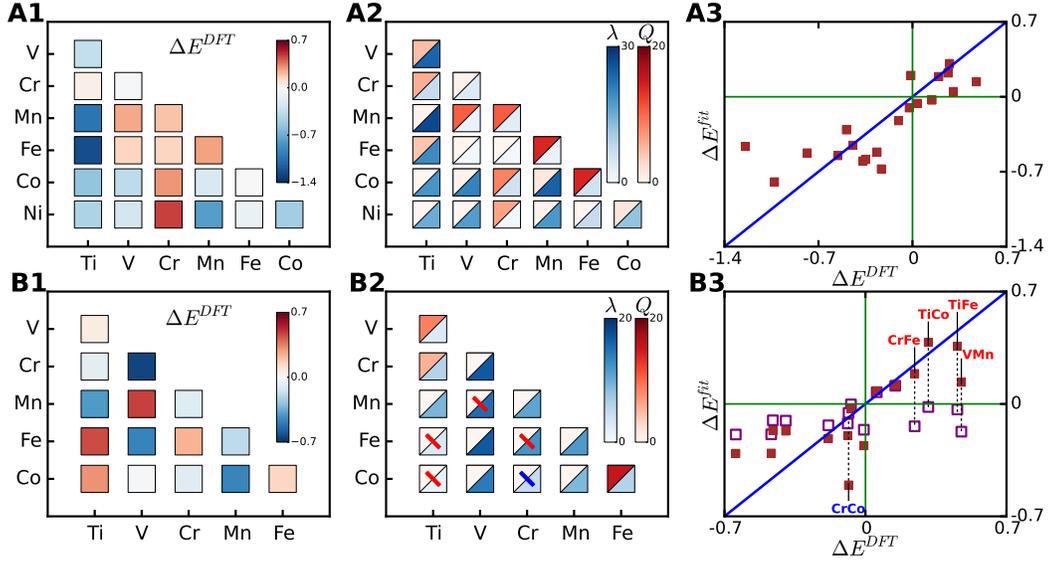}
\caption{\label{fig:fit} \textbf{A}: La$M$O$_3$/La$M'$O$_3$. \textbf{A1})
  DFT-calculated defect formation energy $\Delta E^{DFT}$ for each
  combination. The unit is eV per defect.  \textbf{A2}) Structural
  parameters $\lambda$ and $Q$ for each combination with the most
  favorable magnetic structure. $\lambda$ is shown in blue in the
  lower-right triangle and $Q$ is shown in red in the upper-left
  triangle. The unit on the color bar for $\lambda$ and $Q$ are \%.
  \textbf{A3}) Comparison between the fitted defect formation energy
  $\Delta E^{fit}$ and the DFT-calculated defect formation energy
  $\Delta E^{DFT}$. $\Delta E^{fit}$ are obtained by minimizing
  $\Omega$ of Eq.~(\ref{eqn:cost}).  \textbf{B}:
  Sr$M$O$_3$/Sr$M'$O$_3$. \textbf{B1}) Same as \textbf{A1}. \textbf{B2}) Same
  as \textbf{A2}. The red (blue) backslash denotes those combinations
  in which $S_{MM\prime}$ is -1 (+1). The other combinations without a
  backslash has a zero value of $S_{MM\prime}$. \textbf{B3}) Same as
  \textbf{A3}. The solid squares denote $\Delta E^{fit}$ that are
  obtained by minimizing $\Omega_{\textrm{Sr}}$ of
  Eq.~(\ref{eqn:cost2}). The open squares denote $\Delta E^{fit}$ that
  are obtained by minimizing $\Omega$ of Eq.~(\ref{eqn:cost}). The
  combinations with a red or blue slash are explicitly labelled.}
\end{figure*}

We estimate the defect formation energy for a given $MM^\prime$ combination as:
\begin{equation}
\label{eqn:diff} E_{\textrm{formation}}\simeq\Delta
E^{DFT}_{MM^\prime} = E_{MM^\prime}(R) - E_{MM^\prime}(L)
\end{equation}
where $L$ is the configuration of an ideal interface and $R$ is the
configuration of one antisite defect, which for the computational unit
cell used here corresponds to a rocksalt or double perovskite
structure in which the $M$ and $M^\prime$ ions populate alternate unit
cells. We consider both ferromagnetic
(Fig.~\ref{fig:large-cell}\textbf{A1} and \textbf{B1}) and
checkerboard $G$-type antiferromagnetic ordering
(Fig.~\ref{fig:large-cell} \textbf{A2} and \textbf{B2}). For the $L$
configuration, we also test $A$-type antiferromagnetic ordering
(ferromagnetic planes with magnetization alternating between layers),
since this magnetic ordering naturally fits the $L$ configuration
(Fig.~\ref{fig:large-cell} \textbf{B3}). We always select the magnetic
ordering that yields the lowest energy state.

The energetics of antisite defects at oxide interfaces from our
calculations are summarized in Fig.~\ref{fig:fit}\textbf{A1} and
\textbf{B1} (numerical results for energy differences together with
information about the magnetic ordering and whether the system is
metallic or insulating are provided in the Appendix). Blue indicates
negative defect formation energies; for these cases we expect that the
corresponding $MM^\prime$ hetero-interfaces are susceptible to
antisite defects. Red indicates positive defect formation energies,
suggesting those interfaces would be stable against defect formation.

\begin{figure}[t]
\includegraphics[angle=-90,width=0.95\columnwidth]{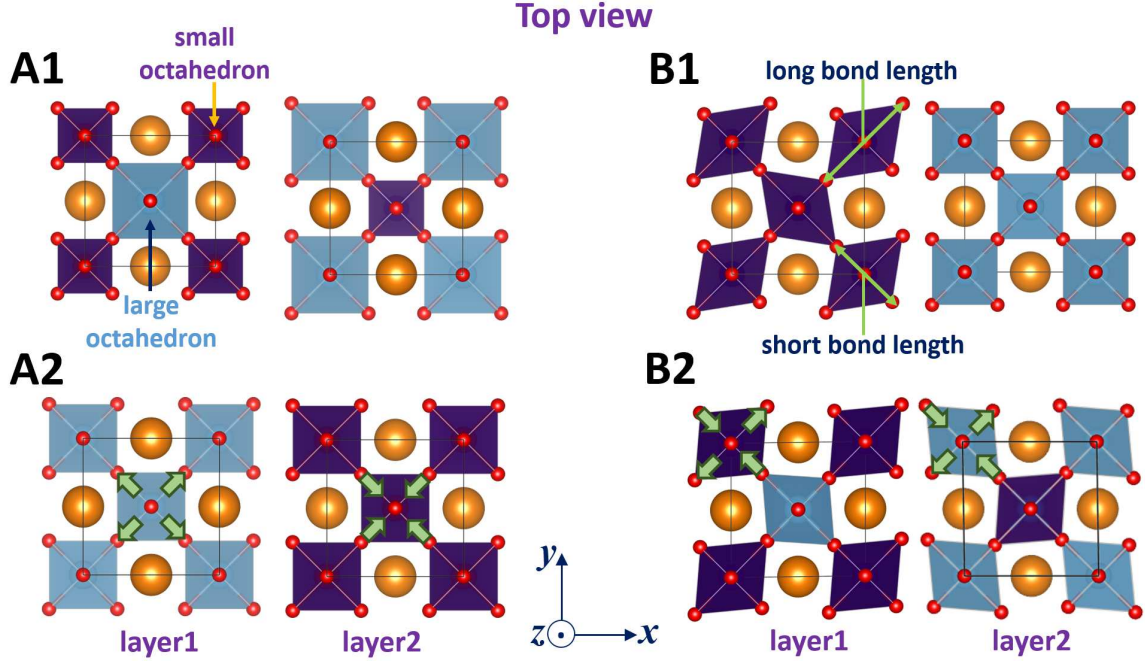}
\caption{\label{fig:mode} \textbf{A1}) Top view of two vertically
  adjacent layers in the $R$ configuration with a large oxygen
  octahedron (blue) and a small oxygen octahedron (purple).
  \textbf{A2}) Top view of two vertically adjacent layers in
  the $L$ configuration with a naturally large oxygen octahedron
  (blue) and a naturally small oxygen octahedron (purple).
  Compatibility with the geometry of the $L$ configuration
  imposes strains (green arrows) on the system, compressing the large
  octahedra and expanding the small ones.  \textbf{B1}) Top view of
  two vertically adjacent layers in the $L$ configuration with one
  anisotropic oxygen octahedron (purple) and one isotropic oxygen
  octahedron (blue). \textbf{B2}) Top view of two vertical
  adjacent layers in the $R$ configuration with one naturally
  anisotropic oxygen octahedron (purple) and one naturally
  isotropic oxygen octahedron (blue). Compatibility with the
  geometry of the $R$ configuration imposes strains (green arrows)
  reducing the bond disproportionation of the anisotropic material and
  inducing a disproportionation in the isotropic one. Rotations and tilts 
  of oxygen octahedra are suppressed for clarity. }
\end{figure}

\section{Local Lattice Distortions and Magnetism\label{Distortions}}

\subsection{Definitions}

In the previous section, we found that for many but not all
$M$/$M^\prime$ combinations, antisite defects were favored. In order
to gain insight into the factors favoring or disfavoring the
appearance of antisite defects, we examine the correlation of the
defect formation energy with other observables.

A basic motif of the perovskite $AM$O$_3$ structure is the volume
$V_M$ of a $M$O$_6$ octahedron. Differences in octahedral volume
between $M$O$_6$ and $M^\prime$O$_6$ octahedra are most easily
accomodated in the $R$ configuration, so we define the $M$O$_6$,
$M^\prime$O$_6$ volume difference as

\begin{equation}
\label{eqn:lambda} \lambda_{MM^\prime} = 2\frac{|V_{M}-V_{M'}|}{V_{M}+V_{M'}}
\end{equation}
with the $V_M$ and $V_{M^\prime}$ evaluated in the $R$ configuration. 

A second important structural variable is a volume preserving
$Q_2$-type Jahn-Teller distortion of octahedron $M$O$_6$ in which one
pair of $M$-O bonds increases in length and the other pair decreases;
both pairs of bonds alternate in the plane (see the left panel of
Fig.~\ref{fig:mode}\textbf{B1}). As will be shown, the mismatch 
of $Q_2$-type Jahn-Teller distortions 
(in-plane octahedral bond disproportionation) between $M$O$_6$
and $M^\prime$O$_6$ octahedra makes an important contribution to
the stability of the $L$ configuration. To quantify this, we define the
bond-length disproportionation for ion $M$ as $Q_M=2|l_1-l_2|/(l_1+l_2)$
where $l_1$ and $l_2$ are the two in-plane $M$-O bond lengths for ion $M$
in the $L$ structure and then define the bond disproportionation
mismatch $Q_{MM^\prime}$ as

\begin{equation}
\label{eqn:Q} Q_{MM^\prime} = |Q_M - Q_{M'}|
\end{equation}
for $M$O$_6$ and $M^\prime$O$_6$ in two adjacent layers in the $L$ 
configuration.

\subsection{Analysis: La-based interfaces}
We begin our analysis with the La$M$O$_3$/La$M'$O$_3$ interfaces,
where for most $MM^\prime$ combinations the ground state is insulating
and the GGA+$U$ method is expected to be reliable.  Results for
$\lambda$ are presented in a color scale in the lower right portion of
the boxes in Fig.~\ref{fig:fit}\textbf{A2}. Comparison to
Fig.~\ref{fig:fit}\textbf{A1} shows that a large octahedral volume
difference is associated with a positive defect formation energy
$\Delta E^{DFT}_{MM^\prime}$. This correlation naturally arises from a
strain effect. Fig.~\ref{fig:mode}\textbf{A1} shows that in the $R$
configuration the large and small oxygen octahedra can be naturally
accommodated. However, the geometry of the $L$ configuration
(Fig.~\ref{fig:mode}\textbf{A2}) requires that the two oxygen
octahedra have equivalent in-plane metal-oxygen bond lengths, inducing
internal strain (represented by green arrows in
Fig.~\ref{fig:mode}\textbf{A2}) relative to the bond lengths preferred
by the given charge configurations, thereby increasing the elastic
energy of the $L$ configuration. We note that although rotations of
oxygen octahedra can accommodate different octahedral volumes, our
calculations on fully relaxed structures show that octahedral
rotations can not reduce enough strain to favor the $L$ configuration.

For bulk La$M$O$_3$ octahedral volumes change $\lesssim 10\%$ (the
variation of lattice constant is $\lesssim 3\%$) as $M$ is varied over
the whole first transition metal row, but larger volume differences
may occur in the superlattices. For example $\lambda=15.1\%$ for
LaTiO$_3$/LaNiO$_3$~\cite{Chen-PRL-2013b} and $\lambda=19.4\%$ for
LaTiO$_3$/LaFeO$_3$ ~\cite{Kleibeuker-PRL-2014}. These very large
$M$O$_6$ volume differences are associated with complete charge
transfer from $M$ to $M^\prime$ ions, i.e. $M^{3+}+M^{\prime 3+}\to
M^{4+}+M^{\prime 2+}$ (La-systems) or $M^{4+}+M^{\prime 4+}\to
M^{5+}+M^{\prime 3+}$ (Sr-systems). The $M$O$_6$ octahedral volume of
electron acceptors expands while that of electron donors
contracts. These substantial charge transfers are driven by large
electronegativity differences between $M$ and $M^\prime$
ions~\cite{Chen-PRL-2013a, Chen-PRL-2013b,Chen-PRB-2014}. A full list
of the combinations in La$M$O$_3$/La$M^\prime$O$_3$ that have a
complete charge transfer from $M$ to $M^\prime$ is given in
Table~\ref{tab:La-combination}. We see that interfaces at which
significant charge transfer occurs are expected to be more susceptible
to antisite defects.

\begin{table}[t]
\caption{\label{tab:La-combination} The combinations in La$_2MM'$O$_6$
  which have a complete charge transfer from $M$ to $M'$ ions (upper
  part) and which have a high-spin $d^4$ configuration (lower part).
  $\Delta E_{MM^\prime}^{DFT}$ is the defect formation energy
  (Eq.~(1) in the main text) using the most favorable magnetic ordering.
  The unit is eV per defect.  $\lambda_{MM^\prime}$ is the $M$O$_6$
  octahedral volume difference using the \textit{R} configuration
  (Eq.~(2)). $Q_M$ describes the magnitude of
  $Q_2$-type Jahn-Teller distortions of ion $M$ using the \textit{L}
  configuration (Eq.~(3)).}
\begin{center}
\begin{tabular}{c|c|c|c}
\hline \hline
\multicolumn{4}{c}{the combinations with a complete charge transfer}\\
\hline
charge configuration  & materials system & $\Delta E_{MM^\prime}^{DFT}$ &
$\lambda_{MM^\prime}$ of the $R$ configuration\\
\hline
$d^1-d^2\to d^0-d^3$ & La$_2$TiVO$_6$ &  -0.35  & $23.9\%$\\
\hline
$d^1-d^4\to d^0-d^5$ & La$_2$TiMnO$_6$ & -1.03  & $27.0\%$\\
\hline
$d^1-d^5\to d^0-d^6$ & La$_2$TiFeO$_6$ & -1.24 & $19.4\%$\\
\hline
$d^1-d^6\to d^0-d^7$ & La$_2$TiCoO$_6$ & -0.56  & $18.4\%$\\
\hline
$d^1-d^7\to d^0-d^8$ & La$_2$TiNiO$_6$ & -0.44  & $15.1\%$\\
\hline
$d^2-d^6\to d^1-d^7$ & La$_2$VCoO$_6$  & -0.37   & $20.4\%$\\
\hline
$d^4-d^6\to d^3-d^7$ & La$_2$MnCoO$_6$  & -0.23   & $24.0\%$\\
\hline
$d^4-d^7\to d^3-d^8$ & La$_2$MnNiO$_6$ & -0.78 &  $18.0\%$\\
\hline\hline
\multicolumn{4}{c}{the combinations with a high-spin $d^4$ configuration}\\
\hline
charge configuration  & materials system & $\Delta E_{MM^\prime}^{DFT}$ &
$Q_M$ of the $L$ configuration\\
\hline $d^2-d^4$    & La$_2$VMnO$_6$ & 0.26 & $Q_{\textrm{Mn}}=11.1\%$, $Q_{\textrm{V}}=0.5\%$ \\
\hline $d^3-d^4$    & La$_2$CrMnO$_6$ & 0.20 & $Q_{\textrm{Mn}}=10.3\%$, $Q_{\textrm{Cr}}=0.0\%$ \\
\hline $d^4-d^5$    & La$_2$MnFeO$_6$ & 0.28 & $Q_{\textrm{Mn}}=14.9\%$, $Q_{\textrm{Fe}}=1.0\%$\\
\hline \hline
\end{tabular}
\end{center}
%\end{sidewaystable}
\end{table}

While charge transfer leads to octahedral volume changes that favor
defects, the mismatch of $Q_2$-type Jahn-Teller distortions between
$M$O$_6$ and $M^\prime$O$_6$ tend to inhibit defects. As can be seen
from Fig.~\ref{fig:mode}\textbf{B2}, if a volume-preserving octahedral
distortion has different amplitudes on the $M$ and $M^\prime$ sites,
it cannot naturally be accommodated in the $R$ configuration (green
arrows indicate strain). However, as Fig.~\ref{fig:mode}\textbf{B1}
shows, as long as the $M$O$_2$ and $M^\prime$O$_2$ sheets have similar
mean in-plane bond lengths, arbitrary layer-dependent $Q_2$-type
Jahn-Teller distortions can be accommodated without strain in the $L$
configuration. $Q_{MM^\prime}$ which mathematically defines the
mismatch of $Q_2$-type distortions in Eq.~(\ref{eqn:Q}), is calculated
for different $MM^\prime$ combinations using the $L$ configuration and
the results for $Q_{MM^\prime}$ are presented in a color scale in the
upper left portion of the boxes in Fig.~\ref{fig:fit}{\bf
  A2}. Comparison to Fig.~\ref{fig:fit}{\bf A1} shows that a large
Jahn-Teller mismatch is associated with a negative defect formation
energy $\Delta E^{DFT}_{MM^\prime}$. Among La-compounds, LaMnO$_3$ has
the strongest $Q_2$-type Jahn-Teller
distortion~\cite{Salamon-RevModPhys-2001}. Our calculations confirm
that the combinations LaVO$_3$/LaMnO$_3$, LaCrO$_3$/LaMnO$_3$ and
LaMnO$_3$/LaFeO$_3$, in which the Mn is in a $d^4$ high-spin state,
all favor the $L$ configuration, thereby tending to suppress antisite
defects. We notice that in all the three cases, $Q_{\textrm{Mn}}$
exceeds 10\%, a value much larger than found for other ions (see
Table~\ref{tab:La-combination}).

%$Q_2$ distortions are driven by Jahn-Teller effects associated with
%orbital ordering in a partly filled $d$-shell. The $e_g$ orbitals
%hybridize most strongly with oxygen so partial occupancy of the
%$e_g$ orbitals typically leads to larger Jahn-Teller effects than
%does partial occupancy of the $t_{2g}$ orbitals.

As with the volume change, the relevant question for the
$Q_2$-typedistortion is the occupancy and spin state in a given
structure, \textit{after} any charge transfer has occurred. Usually
large $Q_2$-type distortions are associated with negligible charge
transfer, because charge transfer tends to create
empty/half-filled/filled $d$ shells which are not Jahn-Teller
active. Examples include LaMnO$_3$/LaNiO$_3$ ($d^4+d^7\to d^3+d^8$)
and LaTiO$_3$/LaMnO$_3$ ($d^1+d^4\to d^0+d^5$). In both cases the
charge transfer moves the Mn configuration away from the high-spin
$d^4$ state that favors Jahn-Teller distortions.

%{\bf I commented out a bit
% of text here, which I dont understand and I dont understand the
% relevant to the paper}.%Similarly thisstability disfavors processes
%that would convert a high-spin $d^3$ or high-spin $d^5$ state to a
%high-spin $d^4$ state. That is, creating a$d^4$ high-spin state in Cr
%ions ($d^3\to d^4$) or in Fe ions ($d^5\to d^4$) via a complete charge
%transfer is unlikely.

The volume difference and Jahn-Teller effects will typically coexist
and compete. To understand how this plays out in practice, we
introduce a cost function

\begin{equation}
\label{eqn:cost} \Omega = \sum_{(MM^\prime)}\left[\Delta E^{DFT}_{MM^\prime} - (\alpha\lambda_{MM^\prime}+ \beta Q_{MM^\prime})\right]^2
\end{equation}
where the sum is over all the combinations
$MM^\prime$. $\lambda_{MM^\prime}$ are calculated using the $R$
configuration and $Q_{MM^\prime}$ using the $L$
configuration. Minimizing Eq.~(\ref{eqn:cost}) yields $\alpha=-3.0$ eV
and $\beta = 2.7$ eV for the La-based heterostructures.  The opposite
signs of $\alpha$ and $\beta$ indicate that the volume change
($\lambda_{MM^\prime}$) and Jahn-Teller effect ($Q_{MM^\prime}$)
compete, as expected. The comparison between $\Delta
E_{MM^\prime}^{DFT}$ and $\Delta
E_{MM^\prime}^{fit}=\alpha\lambda_{MM^\prime}+ \beta Q_{MM^\prime}$ is
shown in Fig.~\ref{fig:fit}\textbf{A3}. While there is non-negligible
scatter, the fit is reasonably good. In particular, the crude model
correctly predicts the stability against defect formation (i.e. the
sign of $\Delta E_{MM^\prime}^{DFT}$) for most cases.

\subsection{Sr-based compounds}

Next we consider the Sr-based compounds. The defect formation energy
$\Delta E_{MM^\prime}^{DFT}$ for Sr$M$O$_3$/Sr$M'$O$_3$ is shown in
Fig.~\ref{fig:fit}\textbf{B1}. The $\lambda_{MM^\prime}$ and
$Q_{MM^\prime}$ for Sr-based heterostructures are calculated and
displayed in Fig.~\ref{fig:fit}\textbf{B2}. As for the La-based
heterostructures, substantial charge transfer leads to large
octahedral volume differences and favors the $R$ configuration
(examples include SrVO$_3$/SrCrO$_3$: $d^1-d^2\to d^0-d^3$ and
$\lambda=5.6\%$; SrVO$_3$/SrFeO$_3$: $d^1-d^4\to d^0-d^5$ and
$\lambda=5.6\%$) while a large mismatch of $Q_2$-type Jahn-Teller
distortions stabilizes the $L$ configuration, for example
SrFeO$_3$/SrCoO$_6$. However, we notice that the combinations
(SrTiO$_3$/SrFeO$_3$, SrTiO$_3$/SrCoO$_3$, SrVO$_3$/SrMnO$_3$ and
SrCrO$_3$/SrFeO$_3$) that strongly favor the $L$ configuration have a
nearly vanishing $Q_2$ mismatch, suggesting the presence of an
additional mechanism in the Sr-based heterostructures.

We believe the additional mechanism acting in the Sr-based
heterostructures is itinerant ferromagnetism, which if present in the
$L$ configuration but not in the $R$ configuration, stabilizes the $L$
configuration. The $MM^\prime$ combinations with a
ferromagnetic-metallic ground state in the $L$ configuration and a
non-ferromagnetic-metallic ground state in the $R$ configuration are
labelled by a red slash in Fig.~\ref{fig:fit}\textbf{B2}. Comparison
of Fig.~\ref{fig:fit}\textbf{B2} to Fig.~\ref{fig:fit}\textbf{B1}
makes the stabilization effect evident. There is one case
(SrCrO$_3$/SrCoO$_3$) where the $R$ configuration is ferromagnetic
metallic and the $L$ configuration is not. As we will show below, in
this case (labelled by a blue slash in Fig.~\ref{fig:fit}\textbf{B2})
itinerant ferromagnetism does not significantly contribute to the
stabilization of the $R$ configuration.

To model the effects of metallic ferromagnetism, we define
$S_{MM^\prime} = S^{R}_{MM^\prime} - S^L_{MM^\prime}$ where
$S^{R}_{MM^\prime}$ and $S^{L}_{MM^\prime}$ take the value 1 for
ferromagnetic metallic states and 0 otherwise for the $R$ and $L$
configurations, respectively.  We include this term in the cost
function, obtaining

\begin{equation}
\label{eqn:cost2} \Omega_{\textrm{Sr}} = \sum_{(MM^\prime)}\left[\Delta E^{DFT}_{MM^\prime} - (\alpha\lambda_{MM^\prime} + \beta Q_{MM^\prime}+\gamma S_{MM^\prime})\right]^2
\end{equation}
Minimizing the cost $\Omega_{\textrm{Sr}}$ yields $\alpha$ = -1.8 eV,
$\beta$ = 2.0 eV and $\gamma$ = -0.4 eV. We comment that both in
La-compounds and in Sr-compounds, $\alpha$ and $\beta$ are very close
in magnitude and of opposite signs, implying that the physical effects
from the volume mismatch ($\lambda_{MM^\prime}$) and the Jahn-Teller
distortion mismatch ($Q_{MM^\prime}$) are comparable and the
competition between the two effects is a general phenomenon. The fit
with $S_{MM^\prime}$ is shown in Fig.~\ref{fig:fit}\textbf{B3} as
solid symbols and is clearly superior to the fit performed without
$S_{MM^\prime}$ (open symbols). Inclusion of $S_{MM^\prime}$ makes the
$\Delta E^{fit}$ of the combinations with a red slash much closer to
$\Delta E^{DFT}$, indicating a key role of itinerant ferromagnetism in
stabilizing the $L$ configuration. However, for the case with a blue
slash, inclusion of $S_{MM^\prime}$ drives $\Delta E^{fit}$ further
away from $\Delta E^{DFT}$, implying that itinerant ferromagnetism
does not significantly contribute to the stabilization of the $R$
configuration.

We make two comments concerning the Sr-based compounds. First, GGA+$U$
predicts charge ordering in a number of Sr$M$O$_3$/Sr$M^\prime$O$_3$
cases. Two-sublattice charge ordering is compatible with the
$\sqrt{2}\times\sqrt{2}\times2$ computational cell used here. It leads
to an additional contribution to the energy difference that is not
taken into account in Eq.~(\ref{eqn:cost2}). However,
Eq.~(\ref{eqn:cost2}) works reasonably well, indicating that charge
ordering does not substantially affect energetics. A more accurate
many-body method (beyond GGA+$U$) is needed to study the delicate
effects of long-range orderings~\cite{Chan-PRB-2009}. Second,
SrFeO$_3$, which in bulk is in the high-spin $d^4$ configuration, is
not Jahn-Teller active (unlike LaMnO$_3$), presumably due to the
enhanced covalency of Fe-O bonding~\cite{Adler-PRB-2006}.

\section{Conclusion \label{Conclusion}}

In conclusion, we have shown that $M/M^\prime$ antisite defects are
energetically favored in wide classes of $AM$O$_3$/$AM^\prime$O$_3$
heterostructures. The key driver of defect formation is a high degree
of charge transfer across oxide interfaces, leading to large
differences in equilibrium octahedral volume which in turn are most
easily accomodated by antisite defect formation. On the other hand, a
large mismatch of $Q_2$-type Jahn-Teller distortion tends to inhibit
antisite defects due to geometry constraints, as does itinerant
ferromagnetism (in this calculation a signature of coherent metallic
behavior across oxide interfaces). The association of defects with
charge transfer is unfortunate, as charge transfer is an important
route to obtaining new physics~\cite{Chen-PRL-2013a,Chen-PRL-2013b,
  Chen-PRB-2014}.

Experimentally, near-interface `dead layers' of transition metal
oxides are frequently
reported~\cite{Huijben-PRB-2008,Tebano-PRL-2008}, suggesting a
possible relevance of the present calculations to the behavior of real
interfaces. On the other hand high quality LaTiO$_3$/LaNiO$_3$
~\cite{Cao-NatCommun-2016} and LaTiO$_3$/LaFeO$_3$~\cite{Kleibeuker-PRL-2014}
interfaces have been reported. Further experimental studies of
transition metal antisite defects at oxide interfaces would be very
valuable for shedding light on this physics. On the theoretical side,
we note that our calculations are based on the GGA+$U$
approximation. While this method is believed to be a good
approximation to the energetics of insulating systems, further
investigation of selected cases using more sophisticated (but much
more computationally expensive) methods such as dynamical mean field
theory would also be desirable (although we emphasize that getting the
local lattice structure correct is essential) and it is also interesting that
our conclusions for the more metallic Sr compounds might also be revisited
with other methods. We also note that we have used the same $U$-value
for all compounds. This choice is motivated by a desire to investigate
chemical systematics without additional confounding factors but we
note that our experience is that as long as $U$ is not too small
$\gtrsim$ 4 eV and not too large $\lesssim 10$ eV the basic physics of
importance here (charge transfer, octahedral volume, Jahn-Teller
distortions) are not particularly sensitive to $U$.

%We note that at this stage we consider ground-state energies,
%while oxide heterostructures are grown under non-equilibrium
%conditions at high temperatures. It may be that high-temperature
%states are different from ground states or that kinetic barriers
%render the formation of antisite defects less likely. For example,
%high quality LaTiO$_3$/LaNiO$_3$ and LaTiO$_3$/LaFeO$_3$
%superlattices have been
%reported~\cite{Cao-arXiv-2015,Kleibeuker-PRL-2014}, although our
%analysis indicates that antisite defects are energetically favored
%in these systems. Using density functional plus dynamical mean
%field methods to calculate high-temperature states may help
%address this issue. 

%The experimental situation regarding antisite defects is not yet
%clear. Direct spectroscopic studies have not yet been
%reported. Indirect evidence may be found from transport
%measurements. Near-interface `dead layers' are frequently
%reported~\cite{Huijben-PRB-2008, Tebano-PRL-2008}. On the other hand,

\begin{acknowledgments}
H. Chen is supported by the National Science Foundation under grant No. DMR-1120296. A. J. Millis is supported by the Department of Energy under grant No. DOE-ER-046169. Computational facilities are provided via Extreme Science and Engineering Discovery Environment, through award number TG-PHY130003 and via the National Energy Research
Scientific Computing Center.
\end{acknowledgments}

\clearpage
\newpage

\appendix

\section{Energy difference, most favorable magnetic ordering and 
transport properties}

We show in Table~\ref{tab:energy-La} and Table~\ref{tab:energy-Sr} the
DFT-calculated energy differences $\Delta E^{DFT}_{MM^\prime}$,
defined in Eq.~(1) in the main text. Table I is for La-compounds and
Table II is for Sr-compounds. For each combination $MM^\prime$, we
also show the most favorable magnetic ordering and transport
properties for the rocksalt ($R$) and layered ($L$) configurations.
Text in blue (red) indicates those combinations that favor the
rocksalt configuration (layered configuration).

\begin{table}[b!]
\caption{\label{tab:energy-La} La$_2MM'$O$_6$ where $M,M'$ = Ti, V, Cr, Mn, Fe,
Co and Ni. The lower half is the energy difference between the $R$ and $L$ 
configurations $\Delta E =
E(R) - E(L)$. The unit is eV per supercell (20 atoms). The upper 
half shows the most favorable magnetic ordering and transport properties 
for each configuration. `F' means ferromagnetic ordering, `G' means 
$G$-type antiferromagnetic ordering, `A' means $A$-type antiferromagnetic 
ordering. `M' means metallic, `I' means insulating. `CO' means charge ordering.}
\begin{center} 
\scalebox{1.0}{\begin{tabular}{c|c|c|c|c|c|c|c} \hline \hline
  &  LaTiO$_3$ ($d^1$) &  LaVO$_3$ ($d^2$)  & LaCrO$_3$ ($d^3$) & LaMnO$_3$ ($d^4$) & LaFeO$_3$ ($d^5$) &  LaCoO$_3$ ($d^6$) & LaNiO$_3$ ($d^7$)\\
\hline
\multirow{2}{*}{LaTiO$_3$ ($d^1$)} & \multirow{2}{*}{bulk}  & \color{blue}$R$: G-I  &  \color{red}$R$: G-I  & \color{blue} $R$: G-I  & \color{blue} $R$: G-I & \color{blue}$R$: G-I & \color{blue} $R$: G-I \\
                                         &        & \color{blue} $L$: A-I  &  \color{red} $L$: G-I  & \color{blue} $L$: F-M     & \color{blue} $L$: G-I   & \color{blue} $L$: G-I & \color{blue} $L$: G-I \\
\hline
\multirow{2}{*}{LaVO$_3$ ($d^2$)}  &  \multirow{2}{*}{\color{blue}-0.348} &  \multirow{2}{*}{bulk}    & \color{blue} $R$:  G-I  & \color{red} $R$: F-I    & \color{red} $R$: G-I   & \color{blue} $R$: G-I  & \color{blue} $R$: F-I \\
                                          &                     &          & \color{blue}$L$: G-I   & \color{red}$L$: F-I    & \color{red}$L$: G-I   & \color{blue} $L$: G-I  & \color{blue}$L$: G-I \\
\hline
\multirow{2}{*}{LaCrO$_3$ ($d^3$)} & \color{red}\multirow{2}{*}{0.036}  & \color{blue}\multirow{2}{*}{-0.024}  & \multirow{2}{*}{bulk}  & \color{red} $R$: G-I & \color{red} $R$: F-I & \color{red} $R$: F-I  & \color{red} \color{red}$R$: F-M\\
                                   &                     &                      &                        & \color{red} $L$: A-I &  \color{red} $L$: G-I & \color{red} $L$: G-I & \color{red} $L$: F-M\\
\hline
\multirow{2}{*}{LaMnO$_3$ ($d^4$)} & \multirow{2}{*}{\color{blue}-1.028} & \multirow{2}{*}{\color{red}0.264} & \multirow{2}{*}{\color{red}0.196}  & \multirow{2}{*}{bulk}  & \color{red}$R$: G-I & \color{blue}$R$: G-I  & \color{blue}$R$: F-I \\
                                  &                       &                      &                      &                        & \color{red}$L$: G-I & \color{blue}$L$: A-I-CO  & \color{blue}$L$: F-M \\
\hline
\multirow{2}{*}{LaFeO$_3$ ($d^5$)} & \multirow{2}{*}{\color{blue}-1.244} & \color{red}\multirow{2}{*}{0.144}  & \color{red}\multirow{2}{*}{0.144}  &  \multirow{2}{*}{\color{red}0.276}  & \multirow{2}{*}{bulk} & \color{blue}$R$: G-I & \color{blue}$R$: F-M \\
                                   &                      &                      &                       &                      &       & \color{blue}$L$: G-I   & \color{blue} $L$: F-M \\
\hline
\multirow{2}{*}{LaCoO$_3$ ($d^6$)} & \multirow{2}{*}{\color{blue}-0.556} & \multirow{2}{*}{\color{blue}-0.368} &  \multirow{2}{*}{\color{red}0.304} & \multirow{2}{*}{\color{blue}-0.232}  & \multirow{2}{*}{\color{blue}-0.012} &  \multirow{2}{*}{bulk} & \color{blue}$R$: F-I \\
                                  &                        &                      &                      &                       &                      &                         & \color{blue}$L$: G-I \\
\hline
\multirow{2}{*}{LaNiO$_3$ ($d^7$)} & \multirow{2}{*}{\color{blue}-0.444} & \multirow{2}{*}{\color{blue}-0.264} &  \multirow{2}{*}{\color{red}0.476} & \multirow{2}{*}{\color{blue}-0.784} & \multirow{2}{*}{\color{blue}-0.104} & \multirow{2}{*}{\color{blue}-0.492} & \multirow{2}{*}{bulk}\\
                                   &                       &                      &                      &                        &                      &                       &       \\
\hline\hline
\end{tabular}}
\end{center}
\end{table}

\begin{table}
\caption{\label{tab:energy-Sr} Sr$_2MM'$O$_6$ where $M,M'$ = Ti, V,
  Cr, Mn, Fe and Co. The lower half is the energy difference between
  the $R$ and $L$ configurations $\Delta E = E(R) - E(L)$. The unit is
  eV per supercell (20 atoms). The upper half shows the most favorable
  magnetic ordering and transport properties for each
  configuration. `F' means ferromagnetic ordering, `G' means $G$-type
  antiferromagnetic ordering, `A' means $A$-type antiferromagnetic
  ordering. `M' means metallic, `I' means insulating. `CO' means
  charge ordering.}
\begin{center}
\scalebox{1.0}{
\begin{tabular}{c|c|c|c|c|c|c}
\hline \hline
  &  SrTiO$_3$ ($d^0$) &  SrVO$_3$ ($d^1$)  & SrCrO$_3$ ($d^2$) & SrMnO$_3$ ($d^3$) & SrFeO$_3$ ($d^4$) &  SrCoO$_3$ ($d^5$)\\
\hline
\multirow{2}{*}{SrTiO$_3$ ($d^0$)} & \multirow{2}{*}{bulk} & \color{red}$R$: G(I) & \color{blue}$R$: F(I) & \color{blue}$R$: G(I) & \color{red}$R$: G(M) & \color{red} $R$: G(M) \\
&   & \color{red}$L$: F(I)  & \color{blue}$L$: G(I) & \color{blue}$L$: G(I) & \color{red}$L$: F(M) & \color{red}$L$: F(M)\\
\hline
\multirow{2}{*}{SrVO$_3$ ($d^1$)} & \multirow{2}{*}{\color{red}0.056} & \multirow{2}{*}{bulk} & \color{blue}$R$: G(I) & \color{red}$R$: G(M) & \color{blue}$R$: G(I) & \color{blue} $R$: F(I) \\
&  &  & \color{blue}$L$: F-CO(I) &  \color{red}$L$: F-CO(M) &  \color{blue}$L$: A(M) &  \color{blue}$L$: G(M) \\
\hline
\multirow{2}{*}{SrCrO$_3$ ($d^2$)} & \multirow{2}{*}{\color{blue}-0.072} & \multirow{2}{*}{\color{blue}-0.644} & \multirow{2}{*}{bulk} & \color{blue}$R$: G(M) & \color{red} $R$: G-CO(M) & \color{blue}$R$: F(M) \\
&  &   &  &  \color{blue}$L$: G-CO(M) &  \color{red}$L$: F(M) &  \color{blue}$L$: A-CO(M) \\
\hline
\multirow{2}{*}{SrMnO$_3$ ($d^3$)} & \multirow{2}{*}{\color{blue}-0.396} & \multirow{2}{*}{\color{red}0.476} & \multirow{2}{*}{\color{blue}-0.088} & \multirow{2}{*}{bulk} & \color{blue}$R$: F(M) & \color{blue}$R$: F(M) \\
&  &   &   &  &  \color{blue}$L$: F(M) &  \color{blue}$L$: F(M) \\
\hline
\multirow{2}{*}{SrFeO$_3$ ($d^4$)} & \multirow{2}{*}{\color{red}0.456}  & \multirow{2}{*}{\color{blue}-0.468} & \multirow{2}{*}{\color{red}0.244} & \multirow{2}{*}{\color{blue}-0.184} & \multirow{2}{*}{bulk} & \color{red} $R$: F(M)\\
&  &   &   &   &   & \color{red} $L$: F(M)\\
\hline
\multirow{2}{*}{SrCoO$_3$ ($d^5$)} & \multirow{2}{*}{\color{red}0.312} & \multirow{2}{*}{\color{blue}-0.008} & \multirow{2}{*}{\color{blue}-0.084} & \multirow{2}{*}{\color{blue}-0.460} & \multirow{2}{*}{\color{red}0.268} & \multirow{2}{*}{bulk} \\ &   &   &   &    &   & \\
\hline\hline
\end{tabular}}
\end{center}
\end{table}

\newpage
\clearpage

%\bibliography{anti-site-v22}

\end{document}